# SCIENTIFIC RELEVANCE AND FUTURE OF DIGITAL IMMORTALITY AND VIRTUAL HUMANS

## Cebo Daniel
### *Independent Investigator, Berlin, Germany*

## ABSTRACT

*We are on the threshold of a significant change in the way we view digital life, which will have a major effect on the physical world. Computers have increasingly emulated deceased human beings through growing awareness in the fields of artificial intelligence, big data, and machine learning, and have symbolically managed to overcome death with the help of technology. One thing is clear, though: now that there are proper and legitimate discussions happening about human immortality, we can be certain that the future is upon us. This article attempts to explain and challenge the ways in which digital immortality, in particular, has manifested itself. This paper summarizes the technological solutions, research findings and technical challenges of major researchers by reviewing the key technologies and general technical schemes in the field of digital human beings. The prospects of digital human beings are being investigated.*

## INTRODUCTION

Utilizing virtual assistants like Siri, which gives voice and conversation interfaces, the growth of machine learning techniques for capturing large data sets, and the growing autonomy of computer-controlled systems all represent changes in artificial intelligence that promote the creation of digital immortality. The growth of the acquisition of personality and brain emulation, as well as life after death influenced by computation, will alter the future of religion, affect the perception of the afterlife, and increase the effectiveness of the dead-on society. This paper provides an overview of recent developments in the field of digital immortality, examines how such digital immortals could emerge and raises complicated questions. It presents the first results of a study that has created a virtual persona. This prototype system contains relevant memories, knowledge, processes and modelling of an individual's personality, traits, knowledge, and experience. It is argued this system offers the opportunity for the development of a personality that learns after death [1].

Based on the development plans suggested by the Federation of American Scientists (FAS) and existing hypotheses, from the beginning of the digital human definition to 2014, we have sorted academic achievements and relative references into four aspects called Visible Human, Virtual Physical Human, Virtual Physiological Human and Intelligent Virtual Human. Information on data collection, data processing and data set-up for Visible Person has been obtained. Digital Physical Human references have been outlined in four physical categories, including radiation, ultrasound, electrical, and mechanics. Relative study related to physiological and biochemical changes in the human body has been classified in four ways as gene molecules, cells, and organs for Virtual Physiological Human. Relative study of Virtual Brain Human was primarily based on the virtual human brain and virtual human power.

Digital human simulation blends information technology with bioscience applied to an analogue analysis of DNA molecules and proteins in cells and tissues, as well as organs. It has been widely applied in several fields such as aviation, national defence, film and television, sports and medical care. This paper points out the relative successes and advancement of science in recent years and summarizes the problems and challenges of research practice. This paper aims to provide references for the development of Digital Humans.

## THE IMPACT OF DIGITAL LIFE ON SOCIETY

The consequences of digital life have drastically changed over the last few decades. Without all those used tools, it is scarcely possible to imagine what everyday life will look like. A large proportion of people primarily do their job on a machine, and everyone checks their mobile phones several times every day. Online activities can impact the truth and trust of individuals, through Facebook conversations, etc. Also, the well-being of individuals may be affected both physically and emotionally. Besides, it is possible to affect the well-being of people, both physically and emotionally. In several distinct areas, the effect of digital life on society brings exceptional benefits. Whether it concerns the regulation of a pregnancy or access to knowledge, the healthcare industry profits. Negative consequences are in a particular context alongside those advantages. Digital life can impact relationships, work and learning environments.

The rise of digital life brings with it many negative impacts. All the data can trigger inattentive actions, which in many respects affects personal life, whether it is study or friendship. Information overload is a detrimental consequence of our digital lives. Technology practitioners, scientists, and health professionals surveyed by the Pew Research Center shared their perspectives and findings on individuals from their community. This illustrates that people, no matter what time of the day will feel anxious and worried about immediate access to everything they need. Sleeplessness and impatience are further effects of knowledge overload. A close link to digital life by missing connections with other people may intensify social relationships. A lack of attachment to everyday life is induced by entertainment opportunities and the opportunity to work everywhere.

The effect on friendships is another factor. An investigation by the University of Washington shows that because of their online behaviour, people especially teenagers, have the problem of decreasing offline friendships. Communicating online is simpler for them, and does not involve meetings in person and prevents disputes.

A California State University study demonstrates that technology decreases student attention and induces distraction. It is difficult for such students to hold their focus on one thing for a longer period through behaviours like updating Facebook every 15 minutes. Another phenomenon, induced by multitasking caused by technology, causes negative student output in class.

Human nature will not improve, which is why in the future, violations will still occur. In the future, everything will be open on the Internet, which is why cyber-terrorism will take advantage of that. Privacy, which offers cybercriminals the ability to strike in new ways, would be reduced to a low degree. This will make our world less secure than it is now and personal skills are the only thing that protects people.

As well as being social and being democratic, our concept of being human can change dramatically. That's because the new system is losing relevance and significance. Thus to realize new modes and activities, the world needs to create a new order.

Nevertheless, researchers at the Pew Research Center believe that one of humanity's greatest booms is the Internet. Whether it is schooling, information or family, the opportunity to reach out to other people for all purposes that an individual might think of gives us flexibility and endless choices on how we want to spend our life. This enormous advantage helps us to live our lives today; the internet has made it possible to communicate globally.

Digital life will invent, reinvent and innovate, through networking, private life and career. The Pew Research Center and Elon University research project explored the views of various technology experts, academics and health practitioners on digital life, how life could change in the future, and the well-being that comes with it. Via networking, creativity can be achieved and all the knowledge people can get from it. Such information helps individuals to create new business ideas or simply find an individual job through, for example, linked in by inserting personal information on the website, job offers can be suggested. Careers, but not just careers, can be invented and reinvented.

According to experts surveyed by the Pew Research Center, the healthcare industry obtains enormous benefits through digital growth. Details regarding medical, safety and health services, such as patient medication, can be accessed within seconds. This is a key advantage and has become a game-changer, especially in the areas of elderly care and pregnancy. The quick access to information helps hospitals to identify patients inside the hospital and to provide previous illness information through this the patient can get support more quickly. Without any uncertainties, a pregnant female goes through a controlled process. She knows if the baby is fit and well and if there will be any complications during the pregnancy.

The use of reliable resources is another plus. Digital life helps us to obtain, at any time and almost anywhere, online education. Research data can be easily identified by search engines such as Google, which can provide a single query with thousands of answers. Online platforms allow for a quicker, more convenient way to make purchases from anywhere you want with the added convenience of doing it. The best example is Amazon, for one it is possible to do the buying order more easily. The product range is also much wider than in any stores that may exist. It's easier than ever before to coordinate, arrange and book journeys, vacations or adventures. All the information for the trip and comparison portals can be accessed by interested individuals such as family members, encouraging tenacious searches on various pages and even preventing a drive to a travel agency.

The ongoing growth of knowledge sharing is expected by scientists. Devices are going to have more automation and power, which contributes to the reality that our lives are going to be more and more, and all the choices we make are with cooperating devices. The web will be more integrated but less noticeable as it is now because the web will blend into all our background knowledge.

## MAPPING THE FUTURE OF CLOSED-LOOP BRAIN-MACHINE NEUROTECHNOLOGY

To understand the full potential of brain-machine interfaces (BMIs) designed to communicate between the brain and external devices (such as computers or prostheses), to meet key technological challenges, to develop next-generation closed-loop brain-machine interfaces is needed. Using recorded signals to control devices. The closed-loop BMI continuously records the signals of the brain and nervous system, decoding that information and then encoding the information and sending it to the brain and/or nervous system, usually in the form of local stimuli. The research and development of feedback techniques for the management of neural activity have grown steadily over the past decade, in part due to the prospect that reading and writing in the nervous system are both for the advancement of neuroscience as a field, and therapeutic interventions and related consumer applications [2]. Recently, closed-loop neurotechnology has shown the potential to provide promising therapeutic and rehabilitation opportunities for patients as well as additional opportunities. These efforts include restoring motor function, providing functional treatment for neurological diseases such as epilepsy and Parkinson's disease, treating memory impairments and neuropsychiatric disorders, increasing learning speed and capacity, and devices that can experience and stimulate the activity of our sensorimotor system.

## CLOSED LOOP NEUROTECHNOLOGY

Although the development of closed-loop neurotechnology for BMI is still in its infancy, it is driven in part by the prevalence of neurological and mental disorders, many of which respond poorly to drug therapy or lack other viable treatment options. The advancement of technology shortly will be driven by the capacity to provide proven therapeutic devices for the treatment of chronic depression, post-traumatic stress disorder (PTSD) and diseases such as epilepsy and Parkinson's disease [3]. Loop closure in such devices will provide greater precision and customization as therapeutic stimulation becomes more tailored to respond more directly to the patient's neuronal physiology. Ultimately, next-generation feedback devices will reinvent the partnership between the brain and the body's nervous system, with the potential to provide effective precision electronic medicine and stimulate new applications for consumers in the fields of healthcare, games, and exercise. Given current developmental trajectories, it is likely that next-generation closed-loop neurotechnology that decodes and encodes the neural activity of multiple regions of the nervous system, such as the central nervous system, the peripheral nervous system, and the autonomic nervous system (CNS / SNP / ANS), arise within the next ten to twenty years. To this end, the IEEE Brain Initiative has published a whitepaper that identifies the main challenges and advancements required to successfully develop and implement next-generation closed-loop neurotechnology [4].

Closed-loop technology is the first document to create a technology roadmap outlining the short-term and long-term goals of development cycle G. The guidelines to be followed from this manuscript will direct the implementation of these technologies, with a focus on the efficacy of these technologies in neuroethics and culture, to basic science and clinical translation. The growth of information technology will make digital immortality open to all. The implantation of neuro-nanorobots in the brain will accomplish this. Although they have the same characteristics as normal cells, they differ from their living cousins in many important ways: they are programmed to communicate with computers; They are immortal; They can obtain permits from organic and external sources, etc. These types of nanorobots scan information about a cell, assign it to themselves and then replace the cell. Initially, it is implanted in the blood, then it passes into the vital substance of the biowaste of the organs of the body. The brain exchange is prolonged and painless for the patient, who can continue with their daily activity during and after the procedure. Also, the patient can upload the firmware to improve reality or improve his personality, and after the natural death of the body, the brain sends a signal, it needs to be rationed and connected to a special computer. Thus, the person achieves digital immortality. This brain will be the centre and processor of the data field and, curiously, it will retain its consciousness.

Essentially, it does not matter how well we have prepared these moments. In true sense, most of us are completely blinded by the unintended consequences of our rapid technological development when it comes into motion. Perhaps the most appropriate preparation we can establish is to keep current progress in the world of intelligent technology invaluable and try to think ahead about how changes will shape our future lives - not just socially, But also legally, politically, and philosophically. Where a place in the discussion of digital immortality depends a lot on how one thinks about consciousness [5, 6]. There are two major camps on this question: those for whom consciousness is a product of quantitative processes (materialistic approaches) and those who believe that consciousness represents more than its parts (virtue view). If one would interpret the former, then digital immortality is a question of when and not. This is because scientists will arrive at a stage of technological capability, at which they will artificially imitate the hardware of the brain to preserve their software (the sense of our thoughts, memories, and proposals).

However, when further adjusting we from a Qualia perspective, it does not matter how scientists can imitate the human brain and live there with electrical thoughts and memories. Neuroscientist and psychiatrist Julio Tony to steal a party from the complete zombie "no mass". It is doubtful whether it represents "us". Apart from the issue of consciousness, there were other important philosophical and moral debates about digital immortality, such as whether a person's digital self is entitled after death and whether the digital immortal self can be changed. No, and/or its length is included [7]. The time to save. Still, no matter how exciting the philosophical and ethical debate about digital immortality is, the survival tests done with bleach are certain. The development of new technology by the school is not controlled. Questions and discussions about philosophical or ethical issues are more or less irrelevant in terms of pure innovation and technological progress.

## ONGOING PROJECTS

If some persuasion is needed on this last point, one should just take a look at how many tech companies already want to walk in the door when it comes to extending human life indefinitely. While someone who believes that most of us are alive will already have a huge digital corpse behind us when we die, it's no wonder companies are already looking for ways to monetize this. For viewers of the popular Netflix series Black Mirror, the privatization of our

Digital of Fat life for packaging as a show is a family theme that can help revitalize a loved one. For the rest, it is worth knowing the story of Roman Mazurenko, a Ukrainian entrepreneur whose best friend created a digital version of him using his WhatsApp chat history and social media posts after his death in a car accident. Eugenia Cuida, a co-founder of the Luca company and manufacturer of replicas, the software that 'brought her best friend Mazurenko back to life', did not just drop a curtain that separates life from death [8,9].

Another project on the market today is the Blue Brain Project, a project led by Eckle Polytechnic Fédrell delusion, which seeks to replicate a fully functional human brain. Eternity. I state, "Like a library with people instead of books, forever preserving the memories, ideas, creations and stories of millions of people, or the current interactive history and its interactive. It is a private venture for "explaining history". generation. And he works at the crossroads of digital and biological consciousness in another start-up, Cincinnati Valley in Carnal

If you're wondering where Elon Musk is all this, thanks to his geotechnical engineer Neuralink here, he's working on a "neural race" that can be injected into a kind of mesh sitting on a man. Please be assured that you are out. is. Connect your brain and it to your computer. In a sense, Neuralink is exploring simple digital immortality to the point where the human brain can be enhanced with digital capabilities. However, the result includes a direct biological and technical interface, which inevitably leads to some form of digital preservation. If this still seems unlikely, a physician in the field of Biomac for Knicks will probably read the electrical brain signals of four or fully locked c-in patients and convert them into movement or language. Consider the fact that it is possible. Faced with these questions, the debate over global citizenship seems like a small fry. Concerns about the nationalities and freedoms of travel around us are somehow appearing under the future, the silicone self digitally stored under our robotic world. Still, it is worth considering what the world will look like in a world affected by loneliness / digital immortality, and how it will affect citizenship and freedom of movement [10].

## TECHNOLOGY BEYOND TOMORROW

The leadership of the Future Directions Committee (FDC) considered the increasing worldwide interest in digital transformation at its November 2018 meeting in Vancouver. The decision was taken at the February meeting to initiate a new Digital Transformation project that could benefit from the involvement of many Potential Path Societies/OUs serving as a catalyst.

The digital transformation we are talking about today is largely due to the advancement of technology, i.e., mostly transducers, i.e., sensors and actuators, and semantics extraction equipment, i.e. artificial intelligence auxiliary data. From analytics. The main reason why industries and organizations are interested in global digital transformation, however, is economics. Digital transformation is changing the economy a little bit. The economy of atoms is the economy of scarcity: atoms are limited; If you give an atom, you will not have one. In contrast, the economy of bits is the economy of abundance: you can give bits while maintaining a copy that is no different from the original bits. Besides, the atomic economy has a high transaction cost, i.e., it costs money and resources to move the atoms along the value chain, while the cost of working on bits is very low. It is attracting new, younger players in various industries [12].

Changes in the economic structure caused by digital transformation, in the area of transaction networks and data centres such as support infrastructures, reduce both capital costs (CAPX) for entering a business with bits for management and operating cost (OPEX). Opex is still large (so the number of companies operating in the sector is small). Given the benefits of the bit economy over the nuclear economy, industries are making every effort to move their atom-based operations to the bit domain. Bits and atoms can be integrated by technologies such as augmented reality (AR) and virtual reality (VR). Provides unique access to the world of AR and VR bits and increasingly secures connectivity with physical twins through digital twins.

The digital twin physical entity (such as an object or process or a set of aggregate objects or processes) and the digital shadow of the physical entity both reflect its current state (taken) of monitoring and simulation (and its history). Support for root cause analysis. Digital twins can, in some cases, be used as a proxy for physical twins, which are exploited in Industry 4.0 as well as other areas. AR, in a way, connects the world of bits to atoms, by superimposing bits on atoms. VR, on the other hand, exploits bits and is used, for example, in the design, training or presentation stages of the potential solution to the user. IEEE's current two initiatives have worked on key components of digital transformation: digital reality (AR and VR) and symbiotic autonomous systems (SAS) (digital twins). The results of these two initiatives and communities so far provide an excellent starting point for this new initiative to take advantage of the growing interest of the industry in the exploitation of the bit economy. Numerous industries have already expressed their support, and the most effective way to deliver faster prices is to compile the results of these initiatives [13].

Additionally, ongoing, and past future initiatives and communities will contribute to the new. For example, the IoT initiative covers an important part of digital transformation, and the Future Network Initiative is working on a communication fabric for digital transformation. Wi-Fi 6 is poised to become a commercial reality by the end of 2019, and 5G is emerging from the world of unified telephone communications (802.11) for the first time, simultaneously evolving through IEE-powered computer networking. Capable, capable of uniting. Category quality). The new digital reality initiative, led by Steve Dukes, Roberto Sarko and Raj Tiwari, will support and enable future digital transformation in various IEE communities.

## FUTURE

We have categorized, evaluated, and concluded detailed preference data based on the discrepancies in Digital Human investigations and application criteria in all aspects. In each phase of research, the overall Digital Human Strategy, as well as technological routes, work plans, current states, priorities, outcomes, and implementation value were summarized. In each aspect of Digital Person, we addressed the technological challenges that exist, along with the urgent main technologies, including data collection, data processing, and modelling

## CONCLUSION

Through this project, what has become obvious is that much of the latest software available to build your own digital eternal lack long-term and in-depth expertise to learn and maintain an efficient digital afterlife. We assume, however, that Virtual Barry can be adapted and used to create sustainable digital people and can be retained or removed according to the wishes of those who want to create a pre-death identity or those left behind who want to retain or remove it. The legal problems concerning the management of preservation and privacy concerns, and the legal ramifications of a continuing existence outside the autonomous jurisdiction of the mortal presence, remain both an ethical and legislative conundrum.

Death is complicated in the modern age as we now have posthumous persistence and the possibilities for the physically deceased in ways not conceivable in previous generations to impact contemporary life. Although the development of virtual human beings and the prospects for realistic digital immortality remains some years away, many of these 'opportunities' are both alarming and troubling, along with the further uncertainty of how effectively digital immortals can learn and learn.